\documentclass[journal]{IEEEtran}
\IEEEoverridecommandlockouts                              

\usepackage{graphicx}

\usepackage[font+=footnotesize]{subcaption}
\usepackage[font+=footnotesize,labelfont=bf]{caption}

\usepackage{amssymb,amsmath,xfrac}
\usepackage{color}

\usepackage{fixltx2e}

\usepackage{tocvsec2}
\usepackage{cite}
\usepackage{booktabs} 
\usepackage{bm}
\usepackage{multirow}
\usepackage{gensymb}

\usepackage{epstopdf} 
\usepackage{enumitem}
\usepackage{enumerate}

\usepackage{xfrac}

\usepackage{balance}

\usepackage{url}
\usepackage[hidelinks]{hyperref}

\usepackage{mathrsfs}
\usepackage{balance} 
\usepackage{float}
\DeclareMathAlphabet{\mathbcal}{OMS}{cmsy}{b}{n}

\usepackage[dvipsnames,svgnames,table]{xcolor}
\usepackage[ruled, vlined]{algorithm2e}
\SetArgSty{textnormal}
\SetKwInOut{Input}{Input}
\SetKwInOut{Output}{Output}
\DontPrintSemicolon

\makeatletter
\newcommand{\setalgotoprulecolor}[1]{\colorlet{toprulecolor}{#1}}
\let\old@algocf@pre@ruled\@algocf@pre@ruled 
\renewcommand{\@algocf@pre@ruled}{\textcolor{toprulecolor}{\old@algocf@pre@ruled}}

\newcommand{\setalgobotrulecolor}[1]{\colorlet{bottomrulecolor}{#1}}
\let\old@algocf@post@ruled\@algocf@post@ruled 
\renewcommand{\@algocf@post@ruled}{\textcolor{bottomrulecolor}{\old@algocf@post@ruled}}

\newcommand{\setalgomidrulecolor}[1]{\colorlet{midrulecolor}{#1}}
\renewcommand{\algocf@caption@ruled}{%
	\box\algocf@capbox{\color{midrulecolor}\kern\interspacetitleruled\hrule
		width\algocf@ruledwidth height\algotitleheightrule depth0pt\kern\interspacealgoruled}}
\makeatother

\setalgotoprulecolor{blue!30}
\setalgobotrulecolor{red!30}
\setalgomidrulecolor{green!30}

\begin{document}

\title{A Database of Dorsal Hand Vein Images}%

\author{Felipe~Wilches-Bernal,~\IEEEmembership{Senior Member,~IEEE,}
	Bernardo Núñez-Álvarez,\\
	and~Pedro Vizcaya,~\IEEEmembership{Senior Member,~IEEE}
	
	\thanks{F. Wilches-Bernal is an electrical engineer and researcher \href{mailto:felipewilches@ieee.org}{(felipewilches@ieee.org)}. B. Núñez-Álvarez is is an electrical engineer and tech sales consultant \href{mailto:berna.nu@gmail.com}{(berna.nu@gmail.com)}. P. Vizcaya is with Pontificia Universidad Javeriana, Bogota, Colombia \href{mailto:pvizcaya@javeriana.edu.co}{(pvizcaya@javeriana.edu.co)}. At the moment of the development of this work, all authors were with Pontificia Universidad Javeriana.}
}


\maketitle

\begin{abstract}
The dorsal hand vein has been demonstrated as a useful biometric for identity verification. This work details the procedure taken to collect two databases of dorsal hand veins in a biometric recognition project. The purpose of this work is to serve as a reference for the databases that are being shared with the public. 

\end{abstract}

\section{Introduction}\label{sec:intro}
Biometrics are physical or behavioral characteristics that be used for identity verification~\cite{jain2011introduction}. Hence, these characteristics are expected to be unique for an individual. Commonly used physical biometrics are fingerprints and the iris but other exits such as those relate to the vascular network of the person. Behavioral biometrics include the signature or the gait of an individual. Combined biometrics, that have both physical and behavioral components, such as the voice also exist~\cite{jain2004introduction}. 

The vascular network form due to the fundamental processes of vasculogenesis and angiogenesis and the terminations of these networks are determined to be unique among individuals~\cite{uhl2020handbook}. The vascular network can then be used as a biometric. In practice the vein configuration of: fingers~\cite{shaheed2018systematic, yang2012finger}, hand palms~\cite{rahul2015literature, mirmohamadsadeghi2011palm}, hand dorsum~\cite{tanaka2004biometric,tesis_wilchesnunez,raghavendra2015hand, huang2014hand,liu2020recognition}, wrist~\cite{mohamed2017combining, das2014new, garcia2020vein}, retina in the eye~\cite{islam2012retina, mazumdar2018retina} and sclera in the eye~\cite{das2013sclera}, have all been used as biometrics. 

The dorsal hand vein have been used as a biometric for over two decades~\cite{tanaka2004biometric}. Several studies have been conducted in different countries and even patents have been granted in this regard\cite{ryabov2016portable}. This work documents the effort to collect two databases of dorsal hand vein images. The databases were collected as part of a larger biometric project performed in 2007 and 2008 at Pontificia Universidad Javeriana in Bogota, Colombia~\cite{tesis_wilchesnunez}. The paper describes the image acquisition system used for the image capture and the final databases obtained. The authors of the original biometric work decided to make the databases publicly available in order to foster research in the subject. Novel techniques such as those based on machine learning or artificial intelligence could potentially be used to the shared datasets. The databases can also be used for educational purposes. 

The remaining of this paper is organized as follows. Section~\ref{sec:acquisys} presents the image acquisition system. Section~\ref{sec:veinpics} describes the databases and how they can be found online. Finally, Section~\ref{sec:concl} presents the conclusions and future work. 

\section{Image Acquisition System}\label{sec:acquisys}

This section presents the physical structure developed to capture images with increased visibility of the dorsal veins of the hand.

\begin{figure}[htb]
	\centering
	\includegraphics[clip,trim=0cm 0cm 0cm 0cm,width=0.3\textwidth]{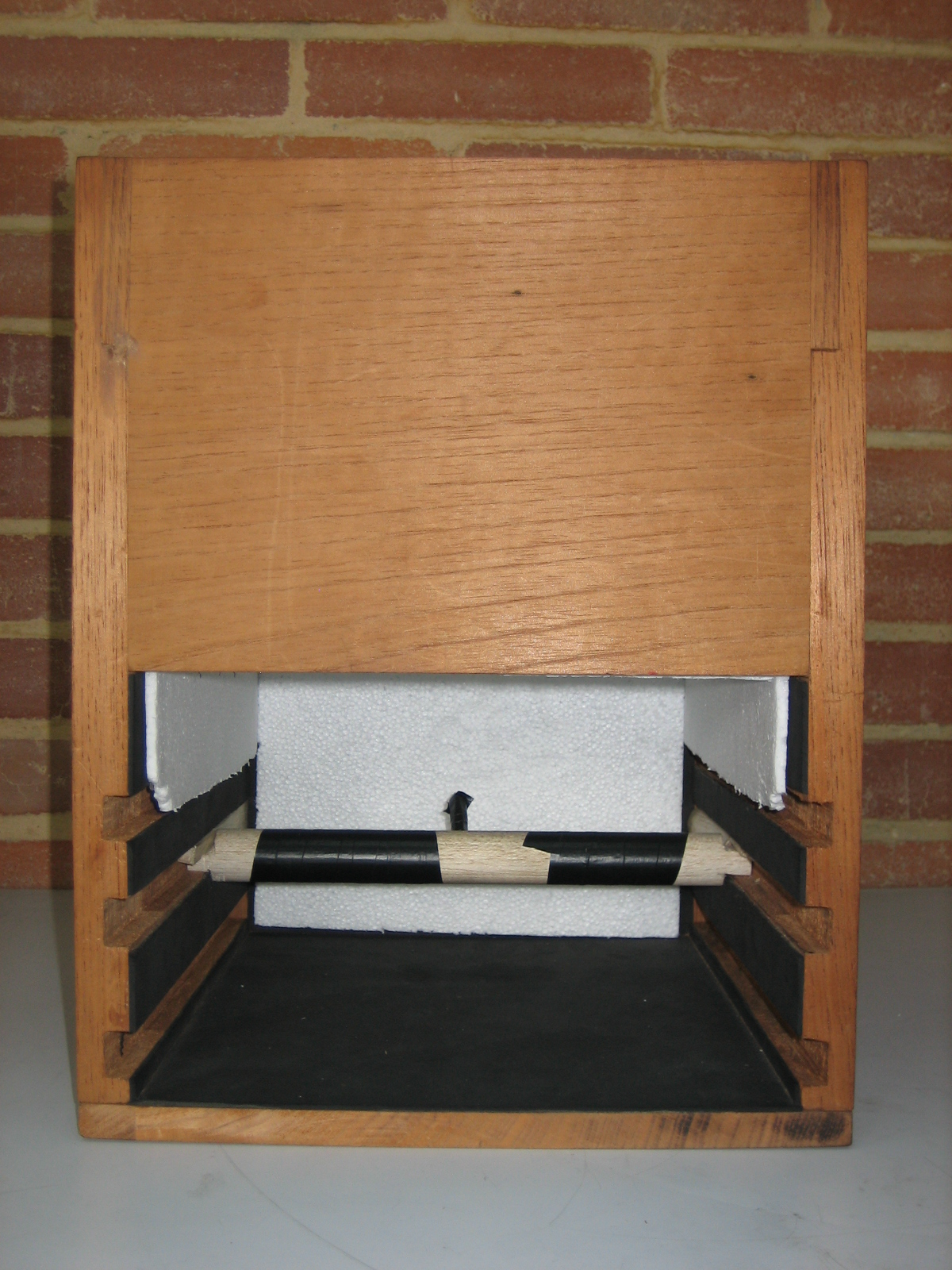}
	\caption{Box-like structure for dorsal hand vein image acquisition.}
	\label{fig:structimacqu}
\end{figure}

\begin{figure}[htb]
	\centering
	\includegraphics[clip,trim=0cm 0cm 0cm 0cm,width=0.3\textwidth]{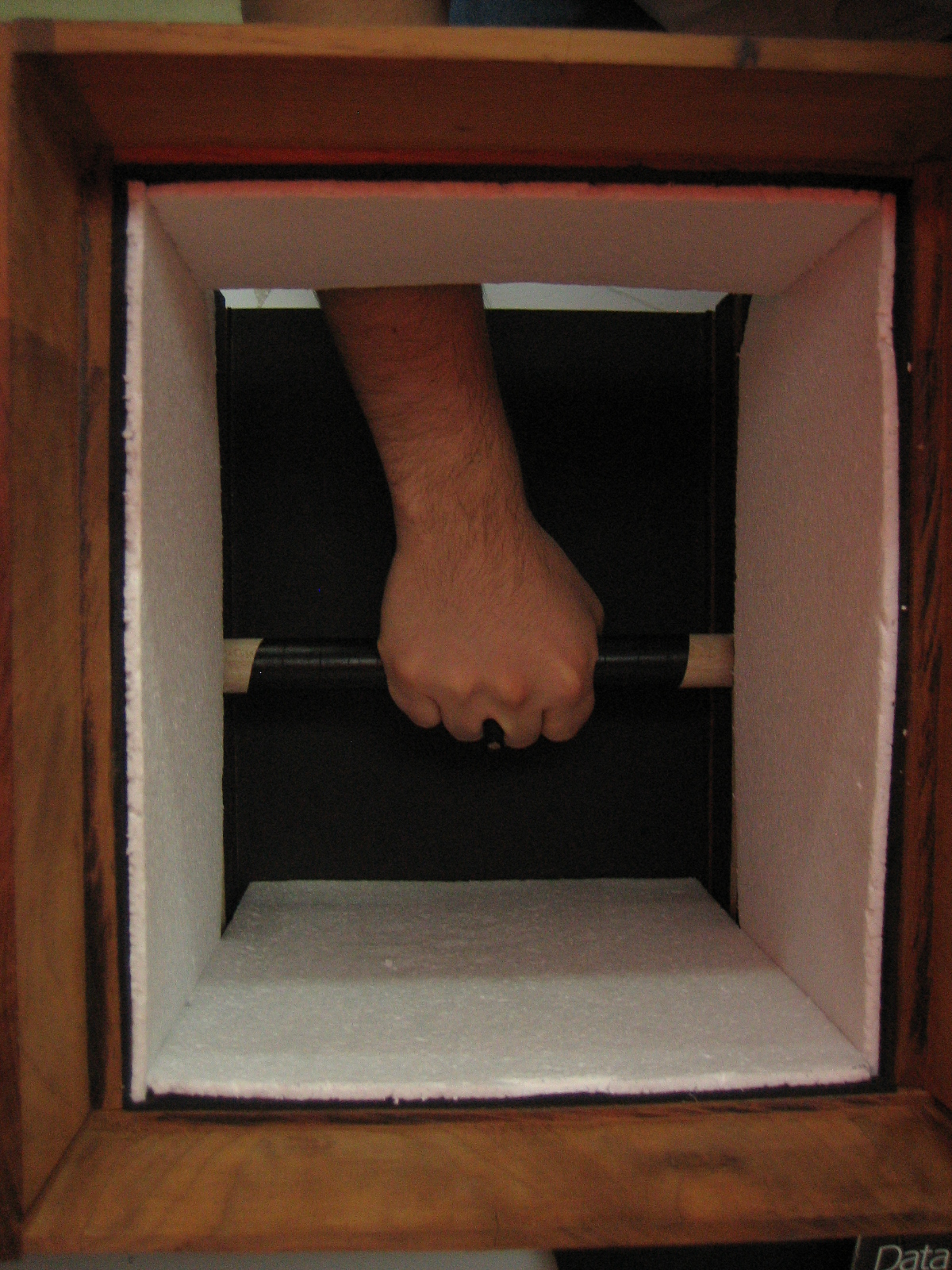}
	\caption{Data acquisition structure: view from above.}
	\label{fig:structimacqu_hand}
\end{figure}

The box-like structure in Fig.~\ref{fig:structimacqu} was built for the purpose of taking dorsal hand vein images. This structure has dimensions of  $32 \times 29 \times 35$ cm  (${\rm L} \times {\rm W}\times {\rm H}$). This structure is built out of wood and its internal walls, except those at the top and bottom, are coated with extruded polystyrene foam (styrofoam) as it can be observed in Fig.~\ref{fig:structimacqu_hand}. The bottom of the structure was covered with black foam and the top of the structure had a custom illumination mechanism and a hole to include the camera, both of which will be explained below.  

The structure also had a rod-line piece that was meant to be grasped by the person whose images were being captured. This rod-like piece had a small stick (or rod) in the middle that was intended to be placed between the middle and ring fingers. Both the rod-like piece and the rod were selected so that the images taken where in a consistent position for different individuals. Note also that these components induce the individual to clench their fist which is a position that tends to increase the visibility of the dorsal hand veins. The box-like structure had different slots where the rod-like structure could be placed. This was done because the distance of the hand to the camera was a parameter of study of the overall project. This study is outside of the scope of this document and all the images collected in the database were at the same distance (of approximately 20 cm).

\subsection{Illumination}\label{sub:illumination}
\begin{figure}[htb]
	\centering
	\includegraphics[clip,trim=0cm 0cm 0cm 0cm,width=0.3\textwidth]{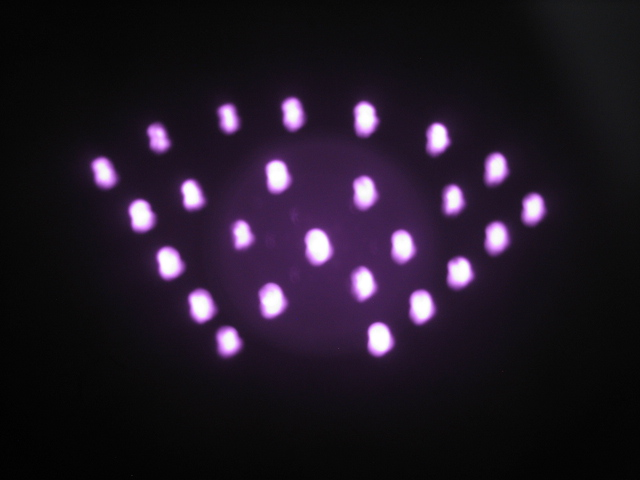}
	\caption{A LED-lamp designed to provide the inner illumination to the data acquisition structure.}
	\label{fig:illum_led}
\end{figure}

The top face of the box-like structure was removable. The inner wall of the top face was equipped with four custom-made LED-lamps such as the one shown in Fig.~\ref{fig:illum_led}. Each one of the four lamps was composed of 25 LEDs for 100 LEDs in total. The LEDs of each lamp were connected in series and supplied with a constant current source. The current sources were designed and controller to provide the same current. Because the light emitted by the LED is dependent in the amount of current, the setup of the project is intended to guarantee that each LED emits roughly the same amount of light. 

The LED lamps were furnished with the QED 223~\cite{qed223} LED device. The QED 223 is an infrared light emitting diode that was selected for the reasons that follow:
\begin{itemize}
	\item The medium wide emission angle has a value of 30$^\circ$ which tend to be a high value for LEDs and is suitable for its purpose of being an illumination source
	\item The wavelength ($\lambda$) of the light emitted has a peak at 880 nm. The wavelength range is between 840 and 950 nm~\cite{qed223}. At these wavelengths the absorption coefficient of the blood in the veins which carry deoxygenated hemoglobin is much higher than that of surrounding tissue.   
	\item The price of this component was low which made it suitable for a self-funded graduation project. 
\end{itemize}

On top of the LED lamps a layer of parchment paper was added in order to scatter the light produced by the lamps. 

\subsection{Camera}\label{sub:camera}
The ISG LW-1.3-S-1394 is the camera used in this project~\cite{camera_lw13s1394}. This device can be connected to a computer using the IEEE-1394 interface. This device was selected because it was already available at the laboratory of the Electronics Engineering Department of Pontificia Universidad Javeriana. Also, this device has an excellent response in the infrared frequency range. In addition this device could be easily configured by software from the computer is connected to. This project used the Image Acquisition Toolbox from MATLAB to configure the camera. The parameters that could be configured were: integration time, frame-rate, gain, brightness, and exposure time. An important feature was that these parameters could be adjusted in real-time from software. Image capture was also easily available from software. 

Because we wanted to capture images only in the infrared spectrum an infrared transmitting filter was added to the top of the lens of the camera. This filter was made from a black processed photographic film (also called a negative).

\subsection{Constrast-enhancing Control System}\label{sub:control}
The project implemented a simple control system to enhance the contrast of the images captured by the camera. The intent of the controller was to increase the contrast of the dorsal hand veins in the captured image as well. The control was also useful to homogenize the color distribution of the image taken for all the people. That is, if the parameters of the acquisition system are kept constant for all people the inherent differences in people's skin colors and their veins would produce images with extremely different contrasts.

\begin{figure}[htb]
	\centering
	\includegraphics[clip,trim=0cm 0cm 0cm 0cm,width=0.45\textwidth]{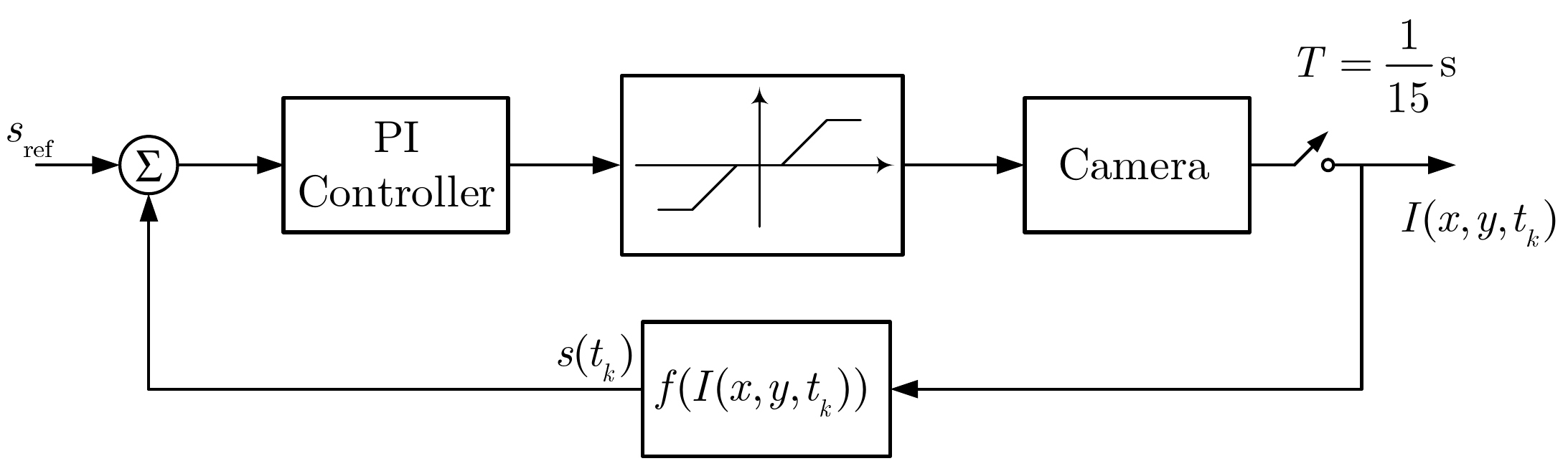}
	\caption{Schematic of the control enhancing system.}
	\label{fig:contcontrol}
\end{figure}

Fig.~\ref{fig:contcontrol} the schematic of the control system implemented to enhance the contrast of the captured image. The variable to control was the camera integration time. Controlling this parameter had a similar effect to controlling the amount of illumination in the hand. The output of the camera is a frame noted $I(x,y,t_k)$ (i.e. a time-varying image or two-dimensional signal). The images have a defined size of $n_r$ rows and $n_c$ columns. The $i^{\rm th}$ row and the $j^{\rm th}$ column of the image are noted, respectively, as $r_i(x,t_k)$ and $c_j(y,t_k)$ and are one-dimensional signals. In order to determine if the frame (image) was suitable for the project, a function to determine the contrast of it was implemented. This function transformed the frame $I(x,y,t_k)$ to a contrast value $s(t_k)$. The function is described by

\begin{equation}\label{eq:Rscal}
s(t_k) = f\left(I(x,y,t_k)\right)= \frac{1}{N_{\rm val}}\sum_{i \in \mathcal{S}} \max (r_i(x,t_k))
\end{equation}
where $\mathcal{S}$ is the set of rows with a maximum above zero\footnote{in practice this meant that the maximum value of the respective column was above a user defined threshold that close to zero (black color).} and $N_{\rm val}$ is the number of elements in $\mathcal{S}$. As seen in Fig.~\ref{fig:contcontrol} the a PI controller was implemented for the task of contrast enhancement. The integral part was selected so the error in steady-state was zero. The reference value $s_{\rm ref}$ was experimentally determined from a set of images that had reasonable contrasts. 


\subsection{Image Capture --Averaging}\label{sub:avging}
The output of the image acquisition system was an image of dorsal hand vein of the individual who was holding the rod. Once the control system in \ref{sub:control} have settled a burst of ten consecutive images (frames) were captured. The final captured image corresponded to the average of those ten frames. An image of variances was also computed and used as a marker to determine whether the captured image was acceptable, this procedure however was not automated. It is important to note that because the frame rate of the camera was set to 15 fps, capturing the ten frames only took two thirds of a second (i.e. 0.6667 seconds). This time was empirically verified as a suitable time for people to hold still the rod.  

\section{Dorsal Hand Vein Image Database}\label{sec:veinpics}
This section presents general information on the two dorsal hand vein image databases generated as part of a biometric project. The project intended to demonstrate that dorsal hand vein images could be used as a reliable biometric for identity verification. This section also presents links to where these images can be retrieved online.

\subsection{Image Collection Procedure}\label{sub:collect}
This project asked people to voluntary participate in data collection sessions.The participants were students and personnel of Pontificia Universidad Javeriana in the year 2007. The ages of the participants vary from 18 to 29 years old. An image collection session was intended to capture several images of the right and left dorsal hand veins. In a session, the participant was asked to grab the rod as shown in Fig.~\ref{fig:structimacqu_hand}. At that point an image was captured as explained in Section~\ref{sub:avging}. Because more than one image per hand was captured per session, the participant was asked to remove the hand from the data acquisition structure, wait for at least 45 seconds, and grab again the rod. This procedure was performed for both the left and right-hands.

The image resolution of the images collected is $752 \times 560$ with 16-bit quantization. The format of the images is \texttt{.tif}. It is important to note that in the year 2020 the images taken in 2007 were parsed and fully anonymized in a Ptyhon algorithm before being rewritten in the same \texttt{.tif} format. These latter images are those shared in the Github repositories. 

\begin{figure}[htb]
	\centering
	\includegraphics[clip,trim=0cm 0cm 0cm 0cm,width=0.45\textwidth]{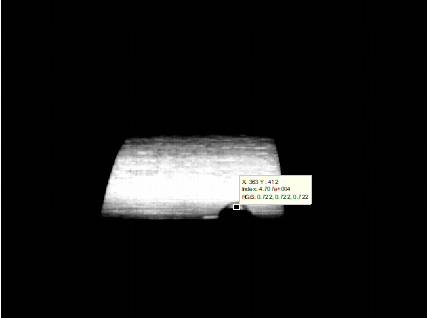}
	\caption{Image captured of the rod.}
	\label{fig:imgstick}
\end{figure}

As explained in Section~\ref{sec:acquisys} the rod to be grabbed by the person had a rod intended to standardize the position of the image. However rotations around the position of the rod were still possible. Fig.~\ref{fig:imgstick} shows the position of the pixel location of the rod in an image without a hand. This location is roughly at row 363 and column 412.

\subsection{Dorsal Hand Veins Image Database 1}\label{sub:db1}
The first database, named Database 1, comprises 138 people, and has 4 images per person per hand for a total of 1,104 images. 

Database 1 can be retrieved at: \texttt{\url{https://github.com/wilchesf/dorsalhandveins}}.

The naming of each image, in Database 1, is as follows:
\begin{itemize}
	\item \texttt{person\_[xxx]\_db1\_[yz].tif} 
\end{itemize}
where \texttt{[xxx]} is a three digit number from 001 to 138 and indicates the number of the person for which the image correspond. \texttt{[y]} is either \texttt{L} or \texttt{R} indicating respectively the left or right hand, and \texttt{[z]} is a number from 1 to 4.

\begin{figure}[h!]
	\centering
	\begin{subfigure}[t]{0.23\textwidth}
		\includegraphics[clip,trim=2cm 0cm 3cm 0cm,width=\textwidth]{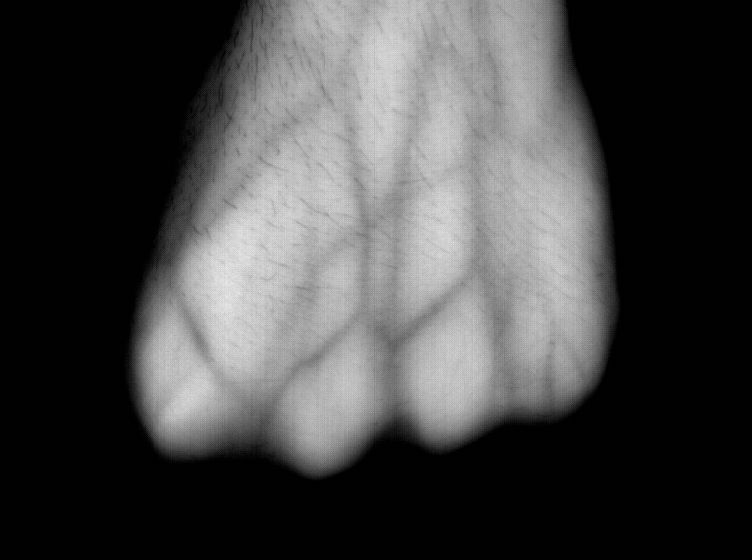}
		\caption{Dorsal left hand vein image of person 18.}
		\label{fig:vp_db1_018_L3}
	\end{subfigure}
	~
	\begin{subfigure}[t]{0.23\textwidth}
		\includegraphics[clip,trim=2cm 0cm 3cm 0cm,width=\textwidth]{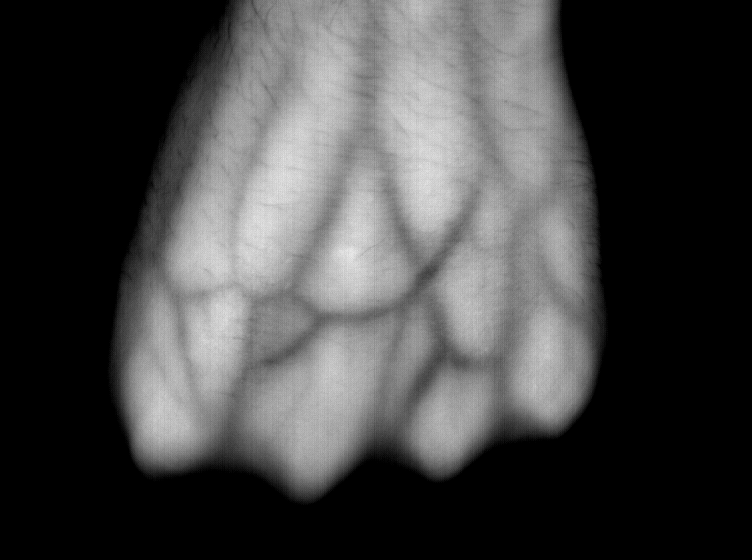}
		\caption{Dorsal left hand vein image of person 54.}
		\label{fig:vp_db1_054_L2}
	\end{subfigure}
	\caption{Image examples of Database 1.}
	\label{fig:vp_db1_ex}
\end{figure}

An example of the images available in the first database is presented in Fig.~\ref{fig:vp_db1_ex}. Note the images where cropped from its original format to fit better the two-column format of this document.

%
%

\subsection{Dorsal Hand Veins Image Database 2}\label{sub:db2}
The second database comprises 113 people, and has 3 images per person per hand for a total of 678 images. All the people in the 113 people in the second database are included in the third database. Due to practical reasons not all the people that participated in the first session of data collection were available for the second session. Each data collection session spanned several days and the the time difference between the data collected in the session for Database 1 and the session for Database 2 is two months.

Database 2 can be retrieved at: \texttt{\url{https://github.com/wilchesf/dorsalhandveins}}.

The naming of each image, in Database 2, is as follows:
\begin{itemize}
	\item \texttt{person\_[xxx]\_db2\_[yz].tif} 
\end{itemize}
where \texttt{[xxx]} is a three digit number from 001 to 113 and indicates the number of the person for which the image correspond. \texttt{[y]} is either \texttt{L} or \texttt{R} indicating respectively the left or right hand, and \texttt{[z]} is a number from 1 to 3.

\begin{figure}[h!]
	\centering
	\begin{subfigure}[t]{0.23\textwidth}
		\includegraphics[clip,trim=2cm 0cm 3cm 0cm,width=\textwidth]{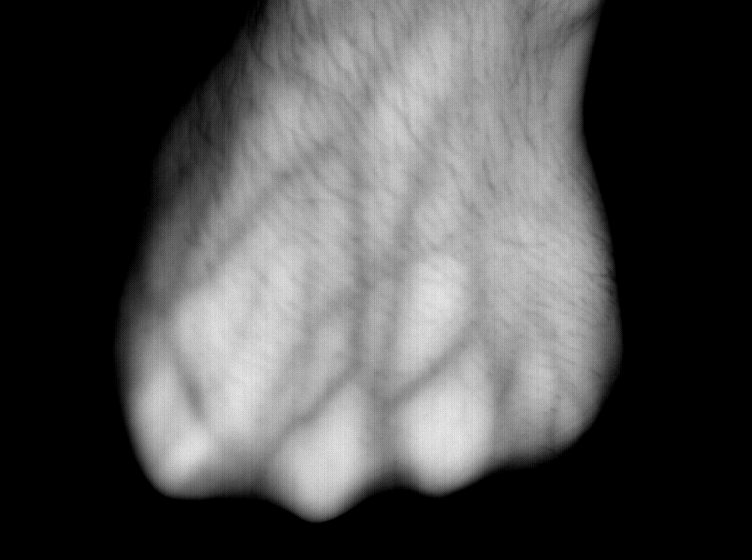}
		\caption{Dorsal left hand vein image of person 18.}
		\label{fig:vp_db2_018_L2}
	\end{subfigure}
	~
	\begin{subfigure}[t]{0.23\textwidth}
		\includegraphics[clip,trim=2cm 0cm 3cm 0cm,width=\textwidth]{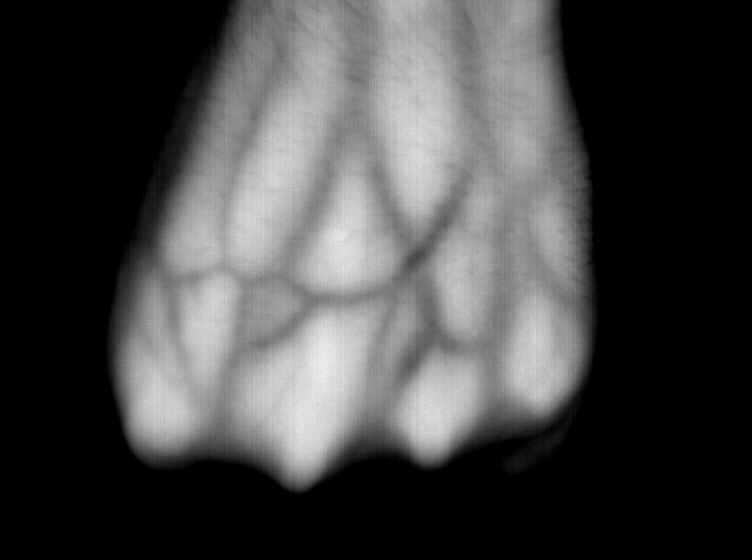}
		\caption{Dorsal left hand vein image of person 54.}
		\label{fig:vp_db2_054_L3}
	\end{subfigure}
	\caption{Image examples of Database 2.}
	\label{fig:vp_db2_ex}
\end{figure}

An example of the images available in Database 2 is presented in Fig.~\ref{fig:vp_db2_ex}. Note the images where cropped from its original format to fit better the two-column format of this document. 

\balance
\section{Conclusions and Future Work}\label{sec:concl}
This work presents two databases of dorsal hand veins that can be used for human recognition. This work details the steps taken to generate the databases. Even though the databases were collected over a decade ago, they are being made publicly available in order to foster research and discussion in the area.  
%
%

\bibliographystyle{IEEEtran}
\bibliography{IEEEabrv,refs_db_veins}
%
\end{document}